\documentclass[twocolumn,preprintnumbers,superscriptaddress,amssymb,prl,aps,10pt,floatfix]{revtex4-1}

\usepackage{graphicx}
\usepackage{dcolumn}
\usepackage{bm}
\usepackage{xcolor}
\usepackage[utf8]{inputenc}
\usepackage[T1]{fontenc}
\usepackage{mathptmx}
\usepackage{natbib}
\usepackage{ulem}

\begin{document}

\preprint{AIP/123-QED}

\title{Densely packed skyrmions stabilized at zero magnetic field by indirect exchange coupling in multilayers}

\author{Fernando~Ajejas}
\email{fajejas@physics.ucsd.edu}
\email{Current address: Department of Physics and Center for Advanced Nanoscience, University of California, San Diego  La Jolla, CA, 92093, USA.}
\author{Yanis~Sassi}
\author{William~Legrand}
\email{Current address: Department of Materials, ETH Zürich, CH-8093
Zürich, Switzerland .}
\author{Titiksha~Srivastava}
\email{Current address: SPEC, CEA, CNRS, Université Paris-Saclay, 91191, Gif-sur-Yvette, France.}
\author{Sophie~Collin}
\author{Aymeric~Vecchiola}
\author{Karim~Bouzehouane}
\author{Nicolas~Reyren}
\author{Vincent~Cros}
\email{vincent.cros@cnrs-thales.fr}
\affiliation{
Unité Mixte de Physique, CNRS, Thales, Université Paris-Saclay, 91767, Palaiseau, France.
}%

\date{\today}

\begin{abstract}
Room-temperature stabilization of skyrmions in magnetic multilayered systems  is the result of a fine balance between different magnetic interactions namely symmetric and antisymmetric exchange, dipolar, perpendicular magnetic anisotropy as well as, in most cases, Zeeman through an applied external field.  
Such field-driven stabilization approach is not compatible with most of the anticipated skyrmion based applications, e.g. skyrmion memories, logic or neuromorphic computing which motivates a reduction or a cancellation of field requirements. 
Here we present a method to stabilize at room-temperature and zero-field, a densely packed skyrmion phase in ferromagnetic multilayers with moderate number of repetitions. 
To this aim, we finely tune the multilayer parameters to stabilize a dense skyrmion phase. Then, relying on the interlayer electronic coupling to an adjacent bias magnetic layer with strong perpendicular magnetic anisotropy and uniform magnetization, we demonstrate the stabilization of sub-60 nm diameter skyrmion at zero-field with adjustable skyrmion density. 

\end{abstract}

\maketitle


Magnetic skyrmions in magnetic heterostructures are non-collinear chiral 2D-like topological spin textures that have attracted great attention in the last decade due to their remarkable properties such as room-temperature (RT) stabilization, small size in the range a few tens of nanometers, current-driven mobility and electrical detection \cite{Fert2013, Everschor2018, Jiang2015, He2017}. Building on a rapid experimental progress, a variety of devices based on skyrmions have been conceptualized for encoding information,  e.g. race track memories, logic devices or neuromorphic computing \cite{Fert2017, Tomasello2014, Finocchio2016, Chauwin2019, Vedmedenko2020, Li2020}. 
In order to control the static and dynamical properties of magnetic skyrmions, different material systems have been investigated, stabilizing either individual skyrmions or skyrmion lattices in ferromagnetic \cite{Romming2013, Jiang2015, Moreau-Luchaire2016,  Boulle2016, Woo2016}, ferrimagnetic \cite{Caretta2018} or more recently 2D materials \cite{wu2020}. In most cases, the application of an external field of at least a few tens of mT is required for its stabilization, as the Zeeman energy assists in transitioning from the topologically trivial maze-domain configuration at zero-field to the skyrmion phase (SP).
Note also that the precise control of the external field allows to finely tune the skyrmion size as well as the density \cite{soumyanarayanan2017, srivastava2021}. On a more fundamental point of view, Büttner {\it et al.} \cite{Buettner2021} recently investigated the formation process of skyrmion lattices at pico-second time scale by combining an ultrashort laser pulse together with a static external magnetic field required to break the time-reversal symmetry. Hence, beyond the fact that the use of external magnetic fields may be an obstacle for the application of skyrmions in new types of computing devices, indeed it is a complication to address experimentally the still-open question to determine how the topological barrier can be overcome.  
No external field requirements shall provide more insights on the topological phase transition mechanism in ferromagnetic (FM) but also in synthetic antiferromagnetic multilayers \cite{Legrand2020, Juge2021}.
To overcome this limitation, stabilization of magnetic textures at zero field seems mandatory. Note that isolated skyrmions or skyrmionic-bubbles in FM multilayers have been already successfully stabilized at zero-field either in confined structures \cite{Boulle2016} or using the interlayer exchange interaction from a perpendicularly magnetized single films \cite{Chen2016, Rana2020}. However, up to our knowledge, there is no experimental evidence of large density of skyrmions or of skyrmion (ordered or disordered) lattice configuration stabilized at zero-field in plain films. A clear advantage working with dense skyrmion ensembles in comparison to isolated skyrmions is for example their interesting applications in reservoir or neuromorphic computing fields \cite{Li2017, Song2020, Grollier2020}.

In this study, we demonstrate that zero external magnetic field (zero-field) skyrmion phase (ZFSP) can be stabilized at RT. To this aim, we first investigate the dependence of the skyrmion size and density on different parameters of the multilayers, with the objective of reducing the field required for the SP
stabilization down to a few tens of mT. Then we describe the approach developed to stabilize ZFSP in multilayers though an electronic indirect exchange interaction generated by an additional bias layer with uniform perpendicular magnetization. The effective magnetic field created by the bias layer is electronically coupled to the skyrmion magnetic multilayers (MML) through a non-magnetic (NM) layer replacing the external field in stabilizing the SP. We demonstrate using magnetic force microscopy (MFM) that by finely tuning the thicknesses of both ferromagnetic (FM) and NM layers, we can obtain ZFSP with apparent skyrmion diameter as small as 60\,nm and an easy control of the skyrmion density.

\begin{figure}[h!]
	\centering
		\includegraphics[scale=0.50]{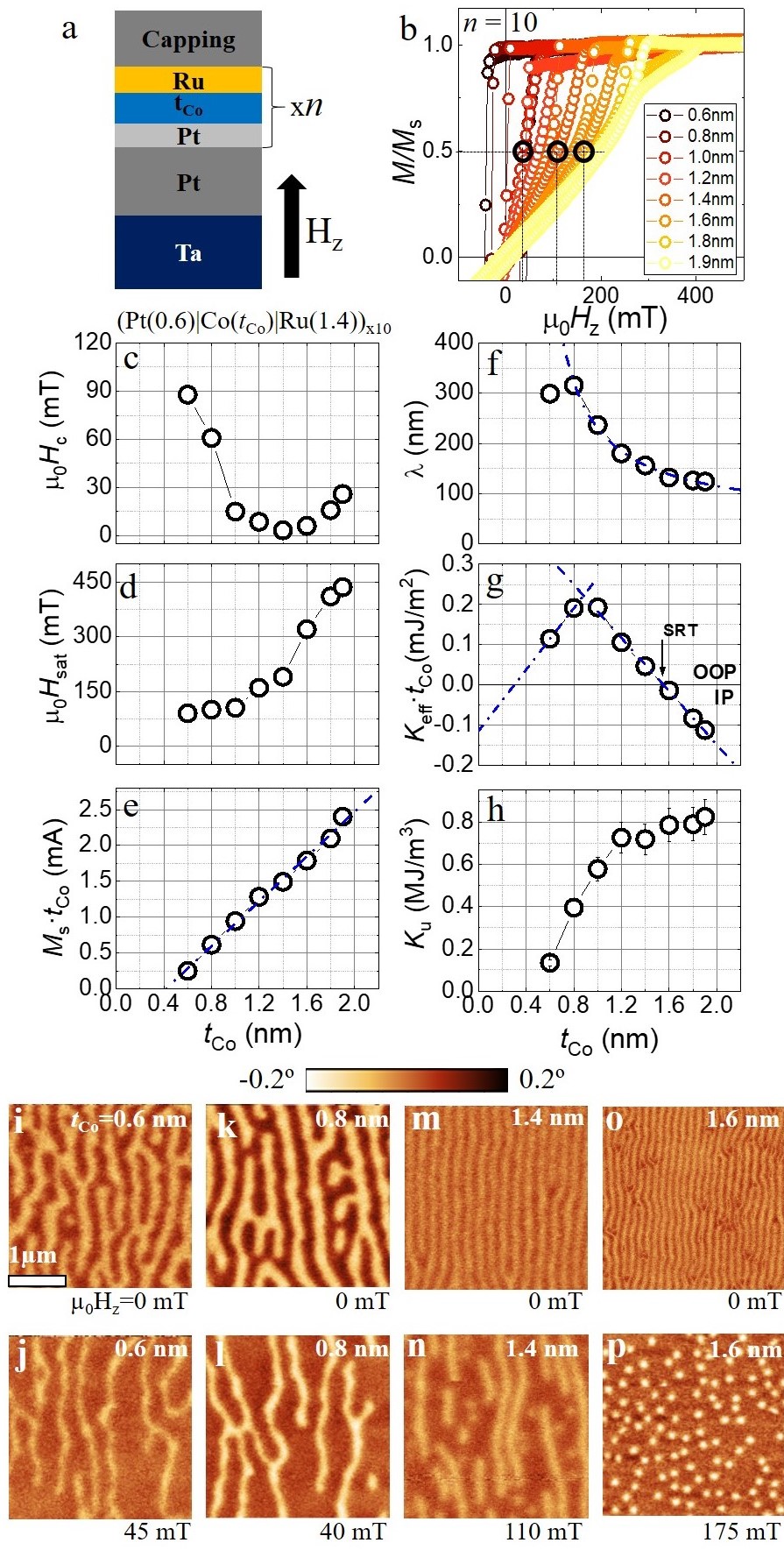}
	\caption{a) Scheme of MML (Pt$|$Co($t_{\rm Co}$)$|$Ru)$_{\times n}$, $t_{\rm Co}$ being the cobalt thickness, $n$ the number of repetitions, and $H_{\rm z}$ the applied perpendicular field. b) Room temperature out-of-plane hysteresis loops of (Pt$|$Co($t_{\rm Co}$)$|$Ru)$_{\times 10}$ for a $t_{\rm Co}$ ranging from 0.6 to 1.9\,nm, measured at room temperature. Black circles indicate the points at which MFM images shown in (j, l, n, p) were taken. c-h) Magnetic parameters as function of Co thickness: c) coercive field $H_{\rm c}$, d) saturation field $H_{\rm sat}$, e) magnetization at saturation $M_{\rm s}$ multiplied by the nominal Co thickness (dotted blue line is a linear fit), f) domain periodicity $\lambda$ after in-plane demagnetization (line described in the text), g) effective anisotropy energy $K_{\rm eff}$ (lines described in the text), and h) calculated uniaxial anisotropy energy $K_{\rm u}$. i-p) Room-temperature MFM images at $H_{\rm z}=0$ and at $H_{z|<M>= M_s/2})$   for $t_{\rm Co}$ $=$ 0.6\,nm (i-j) 0.8\,nm (k-l) 1.4\,nm (m-n) and 1.6\,nm (o-p).}
	\label{fig1}
\end{figure}

The experimental MML are grown by dc magnetron sputtering on thermally oxidized silicon wafers with $280$\,nm of SiO$_{2}$, under $0.25$\,mbar dynamic Ar pressure (base pressure is $7\times 10^{-8}$\,mbar). All the samples are on a buffer made of Ta(5)$|$Pt(8) and capped with 3\,nm Pt layer to prevent oxidation, as schematized in Fig.\,\ref{fig1}(a). Alternating gradient field magnetometer (AGFM) and SQUID are used to measure the anisotropy field $H_{ K}$, and the spontaneous magnetization $M_{\rm s}$. Magnetic imaging using MFM was performed with low-moment magnetic tips in double pass tapping mode-lift mode. A custom made magnetic tip coated with a 7-nm thick CoFeB layer was used for its low magnetic moment in order to limit the perturbation of the magnetic textures in the SP. The MFM setup is equipped with a variable external field module which allows us to modify the external field on demand between two different measurements. The scanned area remains the same regardless small drifts due to the external field and small temperature variations.
The stabilization of the skyrmion configuration and its final characteristics is governed by the balance between the different magnetic energies, i.e., the direct Heisenberg exchange constant ($A$), the Dzyaloshinskii-Moriya interaction (DMI), the magnetic uniaxial anisotropy ($K_{\rm u}$) and the dipolar energies, necessitating a precise magnetic characterization of the samples. 
The effective interfacial DMI ($D_{\rm eff}$) as function of the thickness ($D_{\rm s}$ = $D_{\rm eff}\,t_{\rm Co}$) has been measured by k-resolved BLS to be  $D_{\rm s}$ $=$ -1.27 $\rm pJ/m$ \cite{Ajejas2022, Legrand2022}. We notice that even if the Pt thickness is only $0.6$\,nm it results in an effective perpendicular anisotropy (PMA) and an amplitude of the DMI close to the value of thick layers ($\simeq3$\,nm), as we previously demonstrated in Ref.\,\onlinecite{Legrand2022}. The thickness of Ru of 1.4\,nm leads to a ferromagnetic RKKY interaction between two consecutive Co layers \cite{Legrand2020}.

A schematic view of the studied multilayers composed of a trilayer of (Pt(0.6)$|$${\rm Co}(t_{\rm Co})$$|$Ru(1.4)) repeated $n$ times is shown in Fig.\,1(a) (numbers in parentheses indicate the thickness of each layer in nm). In order to characterize the magnetic properties of the MML as a function of $t_{\rm Co}$, we grew a series of samples with $n=10$ varying $t_{\rm Co}$ from 0.6 to 1.9\,nm. Through the out-of-plane hysteresis loops shown in Fig.\,1(b) together with in-plane hysteresis loops (Fig.\,S1), are determined the magnetic parameters, namely the coercive field ($H_{\rm c}$), the out-of-plane saturation field ($H_{\rm sat}$), the saturation magnetization ($M_{\rm s}$), the effective ($K_{\rm eff}$) and uniaxial anisotropies ($K_{\rm u}$) and the domain periodicity ($\lambda$) measured from the MFM image after an in-plane demagnetization process. Their evolution with $t_{\rm Co}$ are shown in Fig.\,1(c-h). 
$H_{\rm c}$ begins from about 90\,mT at $t_{\rm Co} = 0.6$\,nm,  then decays rapidly reaching a minimum of 3\,mT at $t_{\rm Co} = 1.4$\,nm, then slightly increases (see Fig.\,1(c)). The saturation field $H_{\rm sat}$ displayed in Fig.\,\ref{fig1}(d) shows a continuous increment as a function of $t_{\rm Co}$, with a marked increase above $1.4$\,nm. The evolution of the saturation magnetization $M_{\rm s}\, t_{\rm Co}$ [Fig.\,1(e)] displays a quasi-linear increase with $t_{\rm Co}$. Note that a null magnetization is extrapolated for finite $t_{\rm Co} \approx 0.4$\,nm, suggesting that for this thickness the Curie temperature ($T_c$) is below RT. The slope indicates that the intrinsic $M_{\rm s}$ = 1.56 $\pm$ 0.01\,MA/m, a value close to the bulk Co $M_{\rm s}$ value. In Fig.\,\ref{fig1}(f), we display the domain periodicity after an in-plane demagnetization process, leading to a stripe domain pattern aligned with the external field and a local minimum energy configuration. The observed decrease of the domain period with $\propto 1/t_{\rm eff}$ ($t_{\rm eff} =$ effective magnetic thickness at RT) can be explained by the thickness variation of the total moment ($M_{\rm s}\,t_{\rm Co}$) leading to a lower domain wall energy. 
We then study the evolution of the effective anisotropy $K_{\rm eff}$ extracted from $\mu_{0} H_{\rm sat} = 2K_{\rm eff}/M_{\rm s}$. This value is multiplied by Co thickness ($K_{\rm eff}\,t_{\rm Co}$) to be represented as a function of $t_{\rm Co}$ in Fig.\,1(g). If only shape and interfacial PMA $K_{\rm u,s}$ are considered, then $K_{\rm eff}\,t_{\rm Co} = K_{\rm u,s}-\frac{\mu_0M_s^2}{2}\,t_{\rm Co}$, with $K_{\rm u,s}=K_{\rm u}\,t_{\rm Co}$. 
From $t_{\rm Co}$ = 0.6 to 1.0\,nm, $K_{\rm eff}\, t_{\rm Co}$ first increases reaching the maximum value around $t_{\rm Co}$ = 0.9$-$1\,nm. Thereafter $K_{\rm eff}\,t_{\rm Co}$ decreases linearly up to the largest Co thickness. 
We find the spin reorientation transition (SRT) from out-of-plane to in-plane ($K_{\rm eff}$ = 0) at $t_{\rm Co}$ = 1.53\,nm, in good agreement with other series of (Pt$|$${\rm Co}$$|$Ru)$_{\times n}$ that we studied recently \cite{Legrand2020, Legrand2022}. The linear fit in Fig.1(g) (blue line) is expected to have a slope equal to $-\mu_0M_s^2/2$ in the equation considering the shape anisotropy and a purely interfacial PMA only. This is not compatible with the value of $M_s$ deduced from Fig.1(e), indicating that an additional component is needed, to explain the data in panel (g). Similarly, the calculated $K_{\rm u} = K_{\rm eff}+\mu_0M_s^2/2$ is not constant [Fig.1(h)]. Therefore, we need to consider introducing a magneto-crystalline anisotropy constant $K_{\rm u,c}$. The linear fit in Fig.1(g) suggests $K_{\rm u,c} = -1.9\pm 0.1$\,MJ/m$^3$ and $K_{\rm u,s}=3.8\pm 0.1$\,mJ/m$^2$. 

The changes of the magnetic configuration in these different systems due to the application of an external magnetic field are presented in in Fig.\,1(i-p). The MFM images of the magnetic configuration recorded after in-plane demagnetization correspond to thicknesses $t_{\rm Co}$ = 0.6, 0.8, 1.4 and 1.6\,nm respectively. From these demagnetized states, we then analyze the evolution of the domains as a function of $H_{\rm z}$. The objective is to determine $t_{\rm Co}$ that allows to turn the stripe domains in a densely packed skyrmion configuration at intermediate state in the magnetization loop ($<m_z>\,=\,$0.5). This state is displayed in Fig.\,\ref{fig1}(j) in which the domains pointing opposite to the external field are just shrunken. For moderate but still positive $K_{\rm eff}$ values, isolated skyrmions are stabilized together with meander domains [Fig.\,\ref{fig1}(l)]. The dense SP can be stabilized just before the SRT with slightly positive $K_{\rm eff}$, however, a moderate number of skyrmions combined with domains at $<m_z>\,=\,$0.5 as seen in Fig.\,1(n) are present. Since a larger Zeeman energy is required to reach a SP, this can be only obtained near saturation field. For $t_{\rm Co}$ > SRT $=\,1.53\,\rm nm$ and $K_{\rm eff} < 0$ the stabilization of SP is possible as shown in Fig.\,1(p) in which a SP is stabilized with a density of 10.6\,$\mu$m$^{-2}$.

\begin{figure}
	\centering
		\includegraphics[scale=0.65]{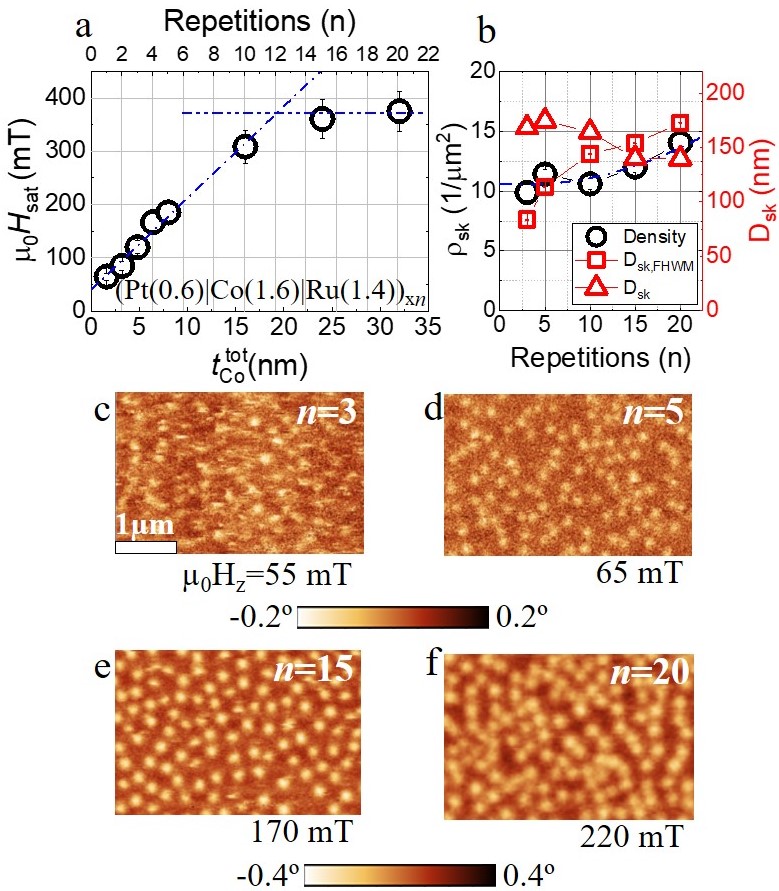}
	\caption{a) $H_{\rm sat}$ as function of the number of repetitions $n$ and total Co thickness ($t_{\rm Co}^{\rm tot}$) of (Pt(0.6)$|$Co(1.6)$|$Ru(1.4))$_{\times n}$ multilayer. The dashed lines are guides for the eyes. b) Skyrmion density, $\rho_{\rm sk}$ (black open circles), and skyrmion apparent diameter $D_{\rm sk}$ extracted from MFM FHWM (red open squares) or calculated from $\rho_{\rm sk}$ and the mean magnetization from AGFM (red open triangles) as function of the number of repetitions $n$. c-f) MFM images of SP of systems with $n=3$ (c), $n=5$ (d), $n=15$ (e), $n=20$ (f). }
	\label{fig2}
\end{figure}

From this analysis, we select $t_{\rm Co}$\,=\,1.6\,nm and $n=10$ that delivers the densest SP at $<m_z> = 0.5$.  
While increasing the number of repetitions increases thermal stability and signal-to-noise ratio, it also implies an increase of the external field that will be needed to stabilize the skyrmions. 
As $<m_z> = 0.5$ is proportional to the saturation field, we study the variation of $H_{\rm sat}$ from the out-of-plane hysteresis loops in different multilayers with $n = 1$ to $n = 20$ repetitions (Fig.\,S2). 
As shown in Fig.\,\ref{fig2}(a), $H_{\rm sat}$ follows approximately a linear evolution up to $n = 12$, above which $H_{\rm sat}$ remains constant. 
In Fig.\,\ref{fig2}(c-f), we display the MFM images as function of the minimum applied external magnetic field $\mu_0 H_{\rm z}$ allowing the observation of the transition from labyrinth configuration into a SP. From the series of MFM images with $n=3$, 5, 10, 15 and 20, we can determine the skyrmion density ($\rho_{\rm sk}$) and the average skyrmion diameter ($D_{\rm sk}$).
The result of the quantitative analysis is presented in Fig.\,\ref{fig2}(b), showing a monotonic increase (40\%) of the $\rho_{\rm sk}$ with the number of repetitions $n$.  
In order to deduce the evolution of $D_{\rm sk}$, we estimate their actual value from $\rho_{\rm sk}$ and the $m_z$ value obtained in the hysteresis loop at the same field that the one used for the MFM images, i.e., $D_{\rm sk} \approx \sqrt{2(1-m_z)/(\pi \rho_{\rm sk})}$ [triangles in right axis of Fig.\,\ref{fig2}(b)]. Following this approach, the $D_{\rm sk}$ is found to decrease by 20\% when $n$ increases from 3 to 20. Another way to estimate the skyrmion diameter is to measure the full width at half maximum (FWHM) of the MFM phase signal corresponding to skyrmions. The results are shown with red squares in Fig.\,\ref{fig2}(b): One finds similar diameter, but the apparent size doubles from $n=3$ to $n=20$. Using this second technique, the apparent size may vary from the actual one due to dipolar and in-plane fields effects, and magnetic tip-sample interactions as well. 
For the rest of the study, we choose $n = 3$, as being the best compromise between the impact of the interlayer dipolar fields and a good thermal stability. Thus, we can stabilize the dense SP with an external field four times smaller than for the sample with $n = 20$.

\begin{figure}
	\centering
		\includegraphics[scale=0.65]{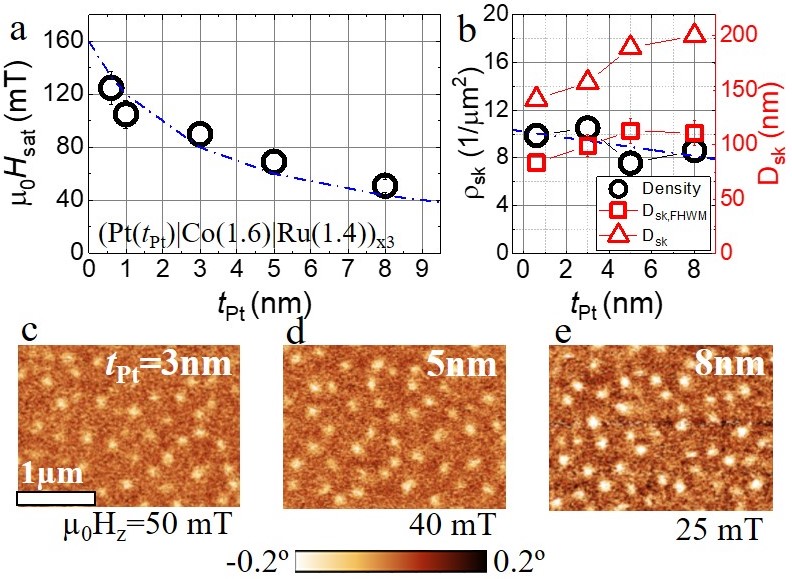}
	\caption{a) $H_{\rm sat}$ vs. Pt thickness ($t_{\rm Pt}$) in $(\rm Pt (t_{\rm Pt})$|$\rm Co(1.6)$|$\rm Ru(1.4))_{x3}$ multilayer. The data is adjusted with 1$/t$ fit.  b) $\rho_{\rm sk}$ (black open circles) and  $D_{\rm sk}$ extracted from MFM FHWM (red open squares) and calculated from $\rho_{\rm sk}$ (red open triangles) as function of $n$, the number of repetitions (n). c-e) MFM images of SP of systems with $t_{\rm Pt}$ = 3\,nm (c), $t_{\rm Pt}$ = 5\,nm (d) and $t_{\rm Pt}$ = 8\,nm (e).}
	\label{fig3}
\end{figure}

For this number of repetitions, $H_{\rm sat}$ can then be further reduced by decreasing the dipolar fields through the increase of the interlayer thickness of the NM layers. As shown Fig.\,\ref{fig3}(a), we find a $1/t_{\rm tot}$ reduction of $H_{\rm sat}$ as function of the Pt thickness ($t_{\rm Pt}$) ranging from 0.6 to 8\,nm, $t_{\rm tot}$ being the total trilayer thickness. In Fig.\,S3 are shown the IP hysteresis loops confirming identical anisotropy values from $t_{\rm Pt}$ = 0.6 to 8\,nm.
In Fig.\,\ref{fig3}(c-e), we present the MFM images showing a SP configuration for $t_{\rm Pt}$ = 3, 5 and 8\,nm. Even though the reduction of the external field is about 60\% between $t_{\rm Pt}$ = 3 and 8\,nm, both $\rho_{\rm sk}$ and $D_{\rm sk,FWHM}$ extracted from the MFM images are found not to change significantly [see Fig.\,\ref{fig3}(b)]. Note however that determining $D_{\rm sk}$ from the $m_z$ and $\rho_{\rm sk}$, the diameter increases by more than 50\,\% when the Pt spacer layer increases from 0.6 to 8 nm [see triangles in Fig.\,\ref{fig3}(b)]. This suggests that the two methods differ more when the dipole fields are smaller. Then, we decide to use the following stacking sequence for the skyrmion MML: (Pt(8)$|$Co(1.6)$|$Ru(1.4))$_{\times 3}$.

\begin{figure}
	\centering
		\includegraphics[scale=0.47]{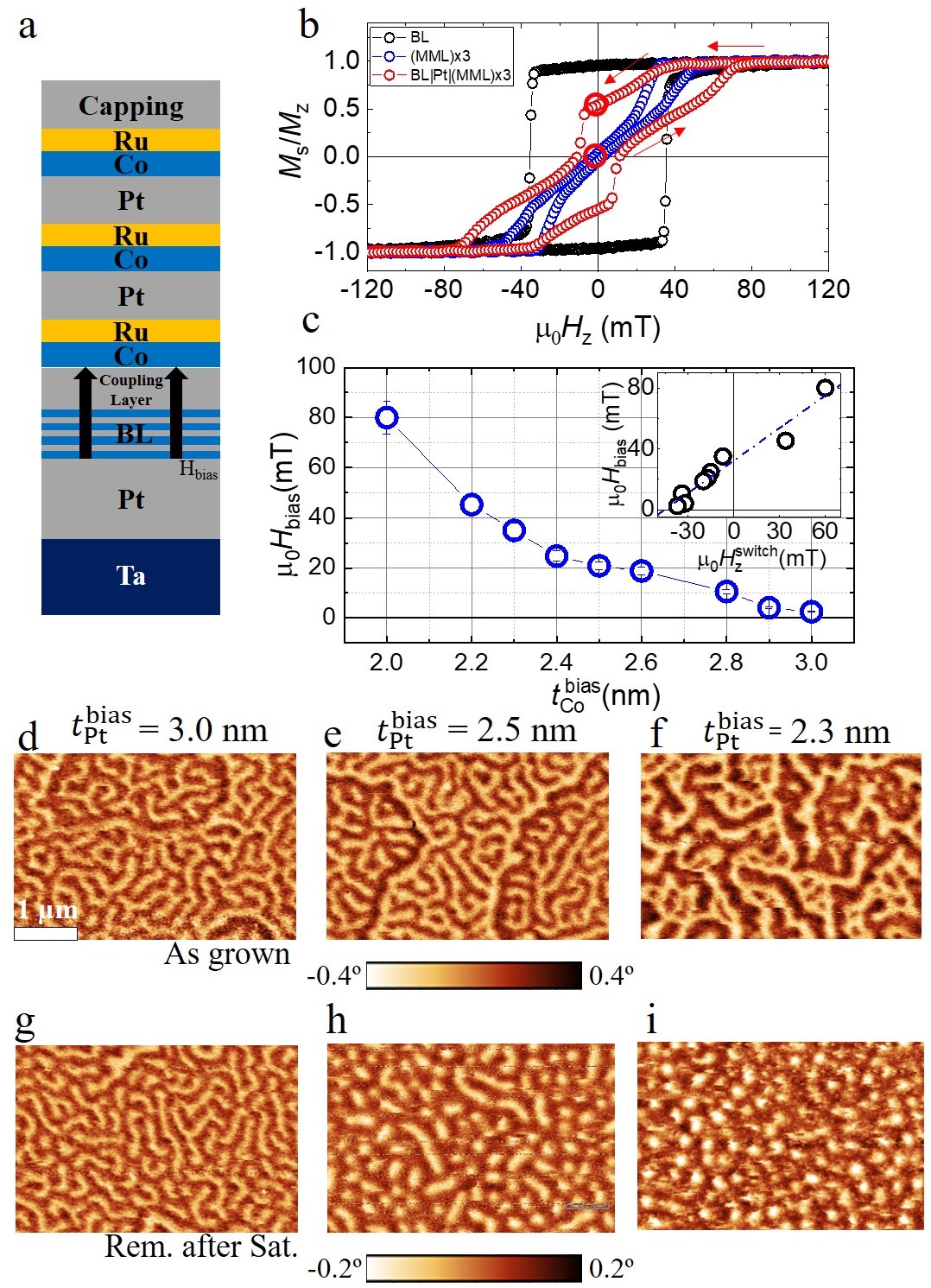}
	\caption{ a) Sketch of ${\rm BL}$|${\rm Pt}(t_{\rm Pt})$|$\rm MML$. BL is the bias-layer with stacking: $({\rm Pt}(0.45)$|${\rm Co}(0.6))_{x4}$, $t_{\rm Pt}$ the Pt coupling-layer thickness and MML the multilayer hosting SP $({\rm Pt}(8)$|${\rm Co}(1.6)$|${\rm Ru}(1.4))_{x3}$. b) AGFM out-of-plane hysteresis loops of BL (black), MML (blue) and $\rm BL$|${\rm Pt(2.3)}$|$\rm MML$ (red). c) Effective bias field ($\mu_{0}H_{\rm bias}$) as function of Pt coupling-layer ($t_{\rm Pt}$). The inset is $\mu_{0}H_{\rm bias}$ vs switching field ($\mu_{0}H_{\rm switch}$). d-i) MFM images of as grown remanence after saturation state for Pt coupling-layer thickness $t_{\rm Pt}$=3.0\,nm (d, g), $t_{\rm Pt}$=2.5\,nm (e, h), $t_{\rm Pt}$=2.3\,nm (f, i). }
	\label{fig4}
\end{figure}

Now that a densely packed SP can be stabilized at low external field values, it can be envisaged to replace the external field by a bias field generated by interlayer electronic coupling with additional layers. In Fig.\,\ref{fig4}(a), a schematic view of the complete sample allowing the zero-field stabilization of SP is presented. In addition to the already designed Pt$|$Co$|$Ru multilayered stack that hosts the skyrmions, it is composed of a bias layer (BL) grown on the the buffer layer (Ta$|$Pt) and a NM spacer Pt coupling layer (CL) through which the indirect exchange coupling is modulated. We show here how the properties of the BL and CL may be engineered. 
We first optimize the BL aiming at reaching a strong enough effective field ($H_{\rm eff} ^{\rm bias}$) needed to stabilize the SP. The first important characteristic of the BL is that it should  have a large PMA together with a completely uniform magnetization at remanence. 
The hysteresis loop of the BL composed of $(\rm Pt(0.4)$|$\rm Co(0.6))_{\times 4}$ is presented as the black open dots curve in Fig.\,\ref{fig4}(b). Note that we have chosen this final composition after having studied the BL properties as a function of the number of repetitions [see Fig.\,S4(a)] showing square shape with sharp transitions. The actual amplitude of $H_{\rm eff} ^{\rm bias}$ is experimentally estimated following the procedure that we developed in Ref. \cite{Legrand2020}. Further information can be found in  Fig. S4(c-d) of the Supplementary Material.
The next step is to optimize the thickness of the Pt CL through which the BL is electronically coupled to the skyrmion MML. We know from the previous section that the $H_{\rm eff}$ amplitude required to stabilize the SP is $\approx\,25\,$mT.
The CL thickness determines the amplitude of the effective bias field $H_{\rm eff} ^{\rm bias}$ acting on the bottom of the skyrmion MML. 
The evolution of $H_{\rm eff} ^{\rm bias}$ as function of $t_{\rm Pt}^{\rm bias}$ coupling-layer is presented in Fig.\,\ref{fig4}(c), in which we see that $H_{\rm eff} ^{\rm bias}$ decays exponentially almost vanishing at $t_{\rm Pt}^{\rm bias}=3$\,nm. Note that the choice of the CL thickness also influences the magnetization reversal process of the complete system. For example, we find that for $t_{\rm Pt}^{\rm bias} < 2.2$\,nm, the effective field  $H_{\rm eff} ^{\rm bias}$ is too large, making that the BL and the skyrmion multilayer reverse simultaneously, hence no skyrmion can be stabilized. 
In the inset of Fig.\,\ref{fig4}(c), we display the experimental results showing $H_{\rm eff}^{\rm bias}$ vs $H_{\rm z}^{\rm switch}$. We define $H_{\rm z}^{\rm switch}$ as the external field applied when the BL  magnetization switching reversal occurs. The interesting coupling regime is when the BL and the skyrmion MML switch independently, keeping a large enough bias field. It corresponds to $t_{\rm Pt}^{\rm bias}$ ranging from $2.2$ to $3.0$\,nm. More details about the reversal mechanisms can be found in Fig.\,S4(b), where the loops are labeled with arrows pointing at $H_{\rm z} ^{\rm switch}$. 
The hysteresis loop of the complete system with a CL of $t_{\rm Pt}^{\rm bias}=2.3$\,nm is displayed with red circles in Fig.\,\ref{fig4}(b), where the blue dot curve is the loop of the skyrmion MML. 
The experimental procedure to prepare magnetically the system leading to the zero-field stabilization of SP consist in first a full saturation of the magnetization of the sample, followed by sweeping the external field back to zero [red arrows in Fig.\,\ref{fig4}(b)]. Red circles are initial and final magnetization state.

The MFM images of the as grown magnetization state and at remanence after saturation for three CL thickness, $t_{\rm Pt}^{\rm bias}$ = 3.0, 2.5 and 2.3\,nm are presented in Fig.\,\ref{fig4}(d-i). First, we see that in the as-grown state, large domains ($\approx \mu$m size) are present in the BL coexisting with smaller labyrinthine domains from the skyrmion MML.
On the contrary, after having saturated the system $H_{\rm z}\geq 80$\,mT and returning to remanence (see Fig.4\,(g-i)), only the magnetic configuration from the skyrmion MML is detected by MFM. For $t_{\rm Pt}^{\rm bias} = 3.0$\,nm, the remnant structure consists of a maze-domain configuration (see Fig.\,\ref{fig4}(g)). For $t_{\rm Pt}^{\rm bias}$ = 2.5\,nm, the magnetic configuration is mainly composed of skyrmions together with elongated domains (see Fig.\,\ref{fig4}(h)). Finally, for $t_{\rm Pt}^{\rm bias} = 2.3$\,nm presented in Fig.\,4(i), the MFM image clearly indicate that the ZFSP is stabilized as only skyrmions stabilized at zero-field are detected.

\begin{figure}
	\centering
		\includegraphics[scale=0.29]{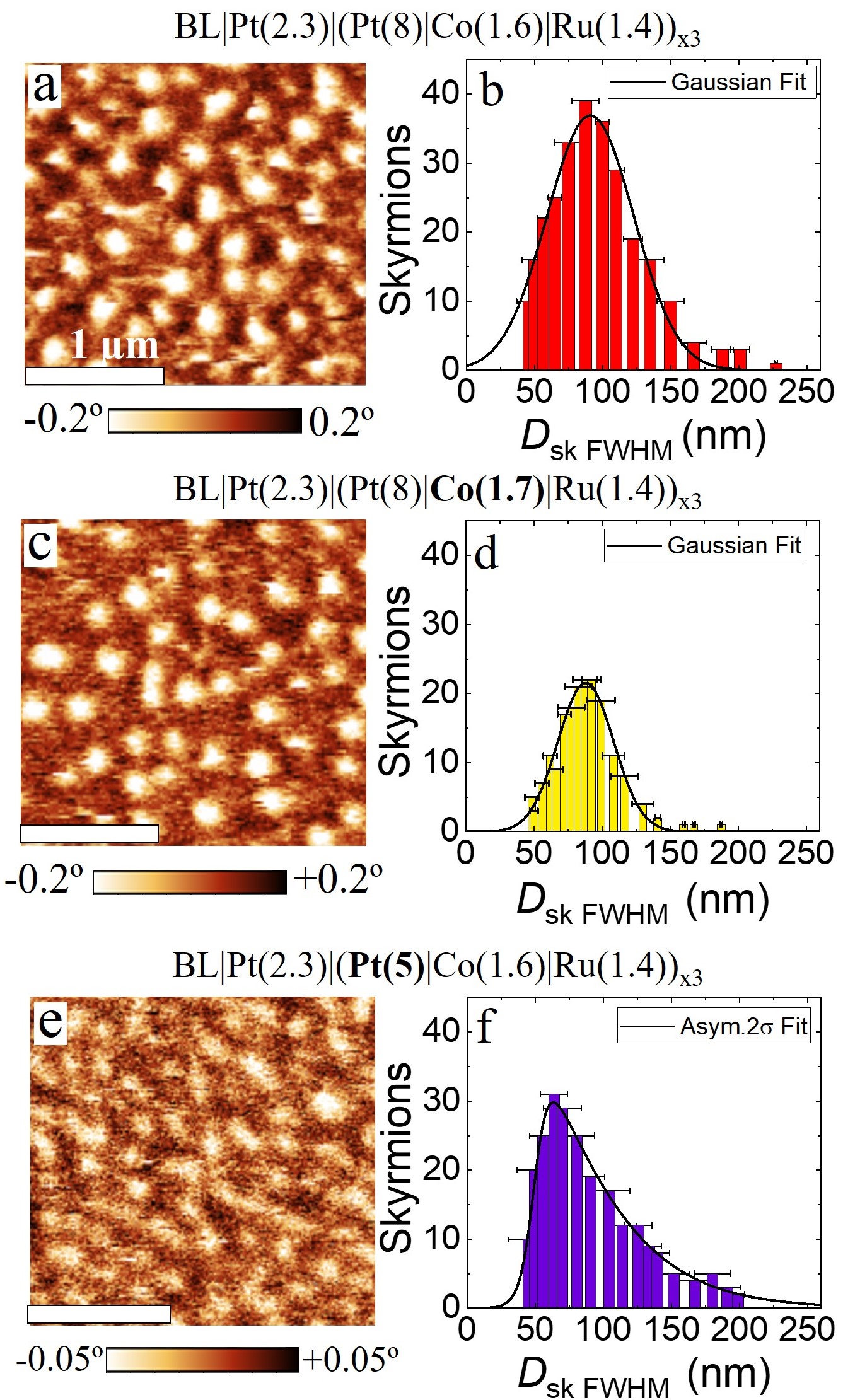}
	\caption{(a) BL$|$Pt(2.3)$|$[Co(1.6)$|$Ru(1.4)$|$Pt(8)]$_{\times 3}$ sample. MFM image of ZFSP at remanence after saturation. (b) Corresponding number of skyrmions distribution as function of $D_{\rm sk,FWHM}$ extracted by MFM image analysis. Black  line is a Gaussian fit. 
	(c) BL$|$Pt(2.3)$|$[Co(1.7)$|$Ru(1.4)$|$Pt(8)]$_{\times 3}$ sample. MFM image of ZFSP at remanence after saturation. (d) Corresponding number of skyrmions distribution as function of $D_{\rm sk,FWHM}$ extracted by MFM image analysis. Black  line is a Gaussian fit.
    (e) BL$|$Pt(2.3)$|$[Co(1.6)$|$Ru(1.4)$|$Pt(5)]$_{\times 3}$ sample. MFM image of ZFSP at remanence after saturation. (f) Corresponding number of skyrmions distribution as function of $D_{\rm sk,FWHM}$ extracted by MFM image analysis. Black  line is an asymmetric Lorentz $2\sigma$ function fit.}
	\label{fig5}
\end{figure}

Finally we investigate the characteristics of the ZFSP, namely $\rho_{\rm sk}$ and $D_{\rm sk}$ that can be tuned by finely varying some of the MML parameters. To this aim, we begin with the optimized system described above, i.e., BL$|$Pt(2.3)$|$[Pt(8)$|$Co(1.6)$|$Ru(1.4)]$_{\times 3}$. Based on this we slightly modify $K_{\rm eff}$ and $H_{\rm eff} ^{\rm dip}$ by:  i) increasing $t_{\rm Co}$ using the system BL$|$Pt(2.3)$|[$Co(1.7)$|$Ru(1.4)$|$Pt(8)]$_{\times 3}$, and ii) by reducing the distance of the FM layers varying the NM bottom layer thickness of the trilayer from 8 to 5\,nm. The experimental system is then BL$|$Pt(2.3)$|$[Co(1.6)$|$Ru(1.4)$|$Pt(5)]$_{\times 3}$ resulting in an increment of $H_{\rm eff}^{\rm dip}$ in the FM layers that will be noticeable.
We first analyze the statistics of the ZFSP of the optimized sample. The corresponding MFM image is shown in Fig.\,\ref{fig5}(a) and the distribution of the number of skyrmions as a function of $D_{\rm sk,FWHM}$ in Fig.\,\ref{fig5}(b). The distribution of diameters can be fitted with a Gaussian function, leading to a mean diameter of $85 \pm 5$\,nm. The density is found equal to $10\,\mu$m$^{-2}$, based on the analysis of a $5\times 5\,\mu$m$^{2}$ MFM image. 

In Fig.\,\ref{fig5}(c), we display the resulting dense SP imaged by MFM at zero field after saturation for the sample with $t_{\rm Co} = 1.7$\,nm. As expected, the correspondingly modified $H_{\rm eff}^{\rm dip}$ leads to energetic variations, which leads to a different ZFSP.
We see in Fig.\,\ref{fig5}(d) that the $D_{\rm sk,FWHM}$ distribution can be fitted with a Gaussian function with mean diameter $\approx 90\pm5$\,nm and with a density of $7.0\,\mu$m$^{-2}$. Note that there is a reduction of 30\% of $\rho_{\rm sk}$ by only increasing $t_{\rm Co}$ by 0.1\,nm (6\% increase). Finally, in Fig.\,\ref{fig5}(e) is shown the MFM image of the system with $t_{\rm Pt} = 5$\,nm. For this sample, we had to use lower magnetization tip to measure without no apparent disruption of the magnetic configuration, leading to a much lower MFM contrast. In this case the $D_{\rm sk,FWHM}$ distribution is fitted with an asymmetric Lorentz $2\sigma$ function presenting a mean diameter $60 \pm 5$\,nm and a density of 12\,$\mu$m$^{-2}$. Obtaining in this case a reduction of the apparent MFM radius of almost 30\%. Note that using 5\,nm thickness of NM bottom layer, the required BL $H_{\rm eff}^{\rm bias}$ is larger than the one for 8\,nm [Fig.3(d-e)], hence there are a few remaining wormy-like domains.

In conclusion, we have investigated how to control the properties of MML to stabilize densely packed skyrmion phase without the need of any external field. Further, we have shown that precisely adjusting magnetic layer properties allows for a fine control of the size and density of the skyrmion phase at zero field with a MML (Pt($t_{\rm Pt}$)$|$Co($t_{\rm Co}$)$|$Ru1.4)$_{\times n}$ with a moderate number of repetitions, {\it i.e.}, $n = 3$ and the larger $t_{\rm Pt} = 8$\,nm. The approach to get the ZFSP is to generate an effective magnetic field created by an uniformly magnetized bias layer that is electronically coupled to the skyrmion MML. We demonstrate that this effective bias field is large enough to transform the as-grown labyrinthine domain configuration into a dense skyrmion phase configuration. We furthermore show that the skyrmion diameter and density can be tuned by slightly modifying the thickness of the trilayer, leading to a variation of the $H_{\rm eff}^{\rm dip}$. We consider that such a precise control of both $\rho_{\rm sk}$ and $D_{\rm sk}$ stabilized at zero-field might allow to facilitate the investigation of the fundamental mechanisms and the time-scale at which the topological barrier can be overcome leading to the nucleation of skyrmion phase. In particular through the use of advanced experimental techniques for which the application of magnetic field is problematic.  Moreover, it also opens some interesting perspectives for the use of skyrmion phase-based systems for neuromorphic computing applications.

This work has been supported by DARPA TEE program grant $(MIPR HR-0011831554)$, by ANR agnecy as part of the “Investissements d’Avenir” program (Labex NanoSaclay, reference: ANR-10-LABX-0035, the FLAG-ERA SographMEM (ANR-15-GRFL-0005) and the Horizon2020 Framework Program of the European Commission, under FETProactive Grant agreement No. 824123 (SKYTOP). 
\bibliography{Zero-field_skyrmionsNotes.bib}

\section{Supplementary Materials}

\begin{figure*}
	\centering
		\includegraphics[scale=0.55]{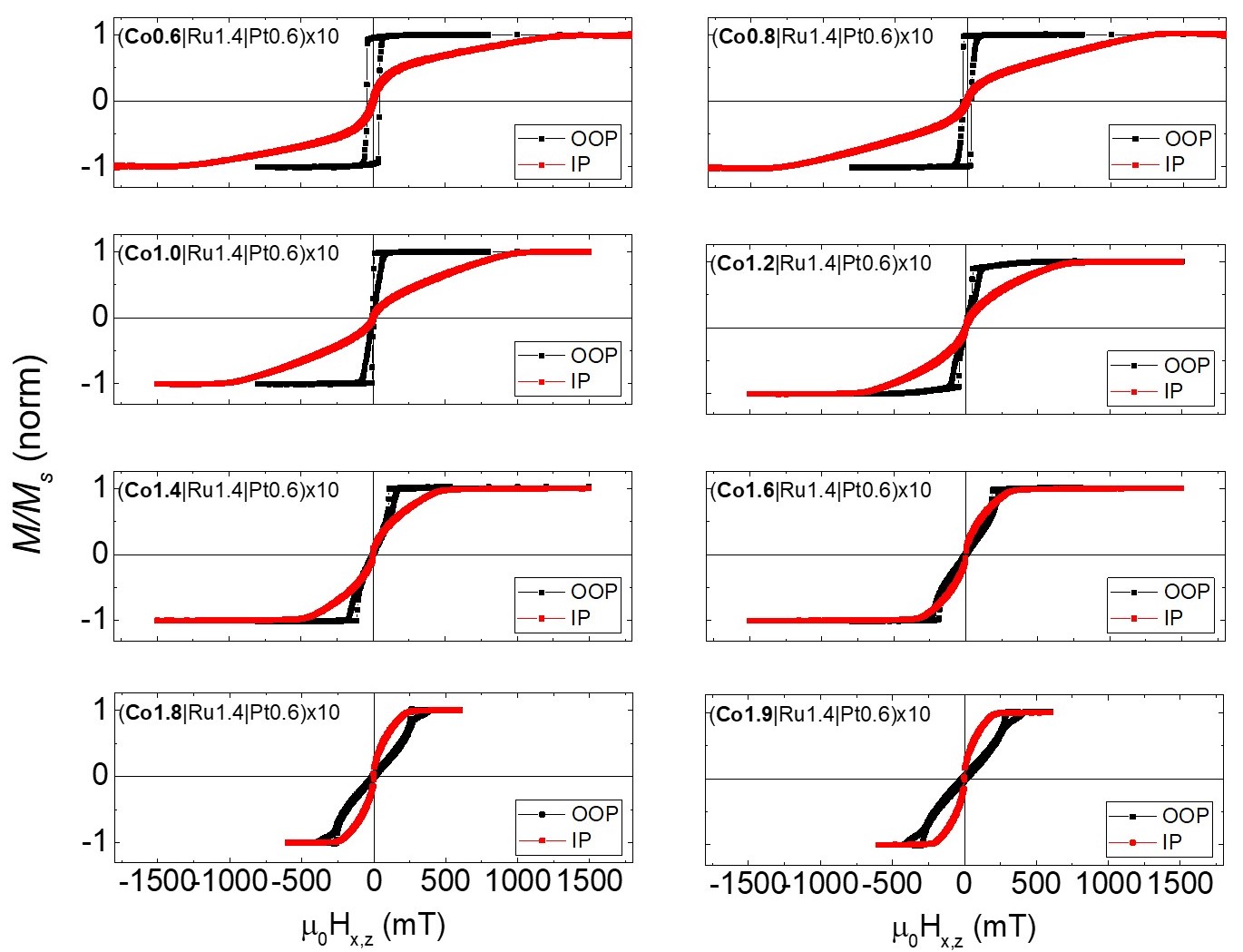}
	\caption{Out-of-plane (black) and in-plane (red) AGFM hysteresis loops of variable $t_{\rm Co}$.}
	\label{figS1}
\end{figure*}

\begin{figure*}
	\centering
		\includegraphics[scale=0.6]{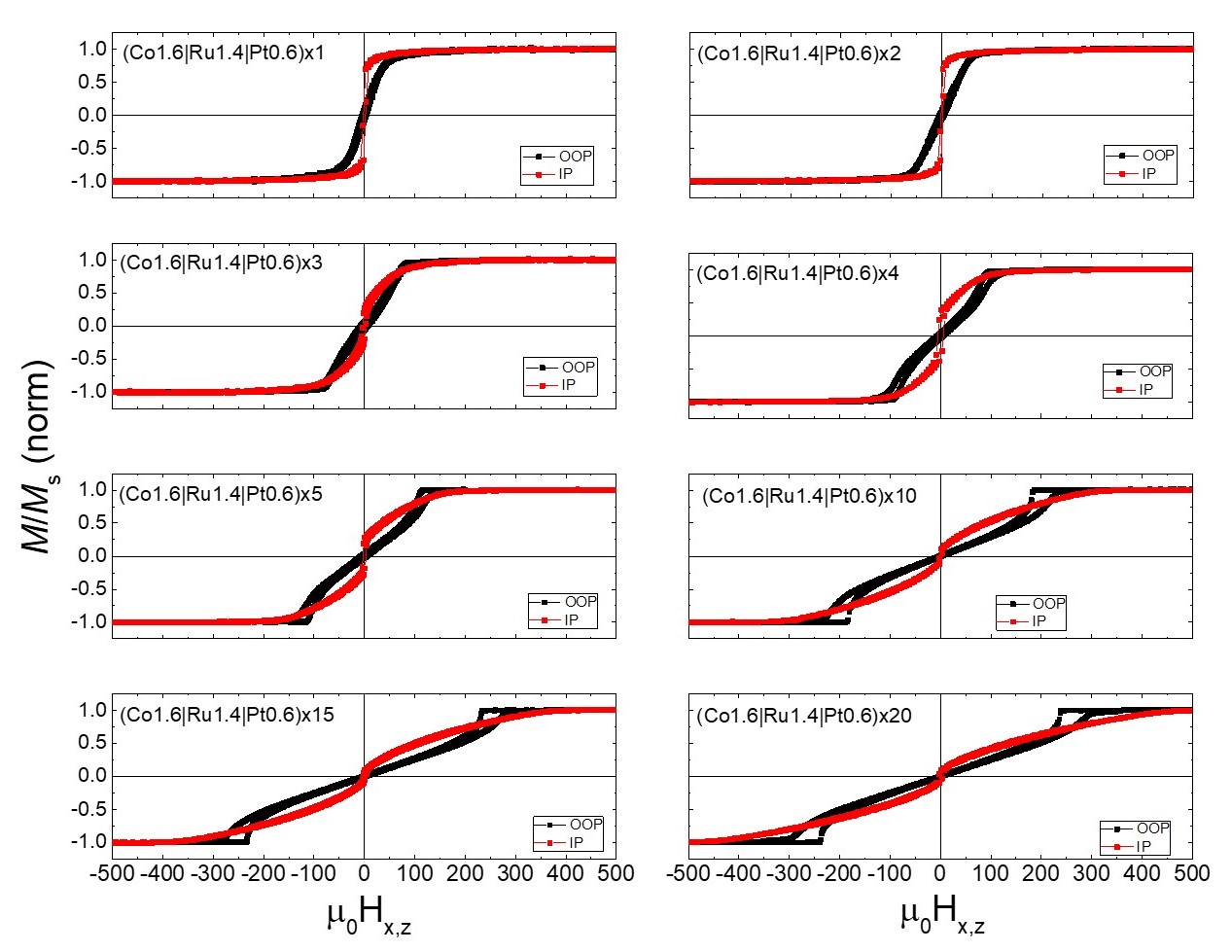}
	\caption{Out-of-plane (black) and in-plane (red) AGFM hysteresis loops of variable $n$}
	\label{figS2}
\end{figure*}

\begin{figure*}
	\centering
		\includegraphics[scale=0.5]{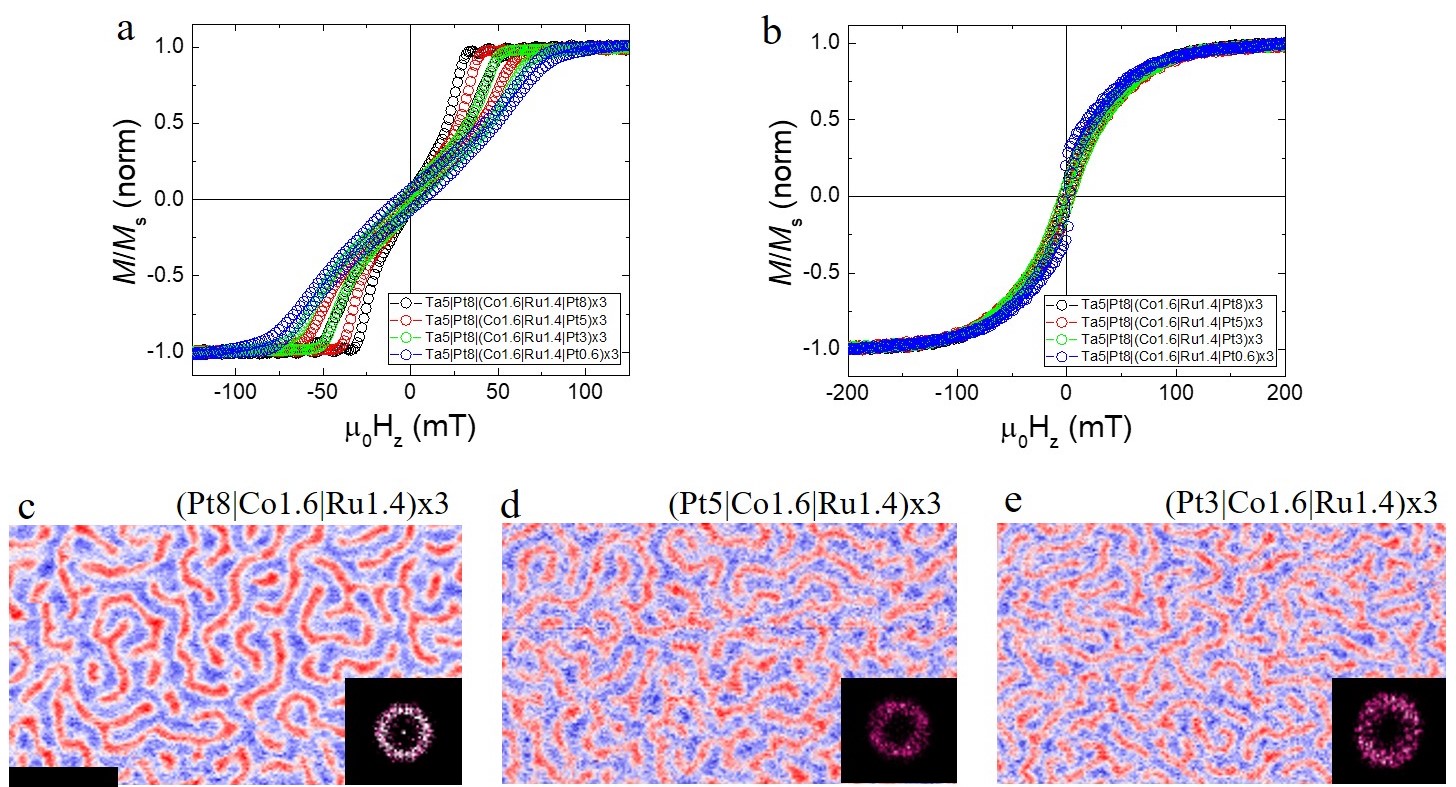}
	\caption{a) Out-of-plane AGFM hysteresis loops of variable bottom NM layer $t_{\rm Pt}$. b) In-plane AGFM hysteresis loops of variable $t_{\rm Pt}$. c) MFM images of OOP demagnetization state Pt(8) d) Pt(5) e) Pt(3).}
	\label{figS3}
\end{figure*}

\begin{figure*}
	\centering
		\includegraphics[scale=0.5]{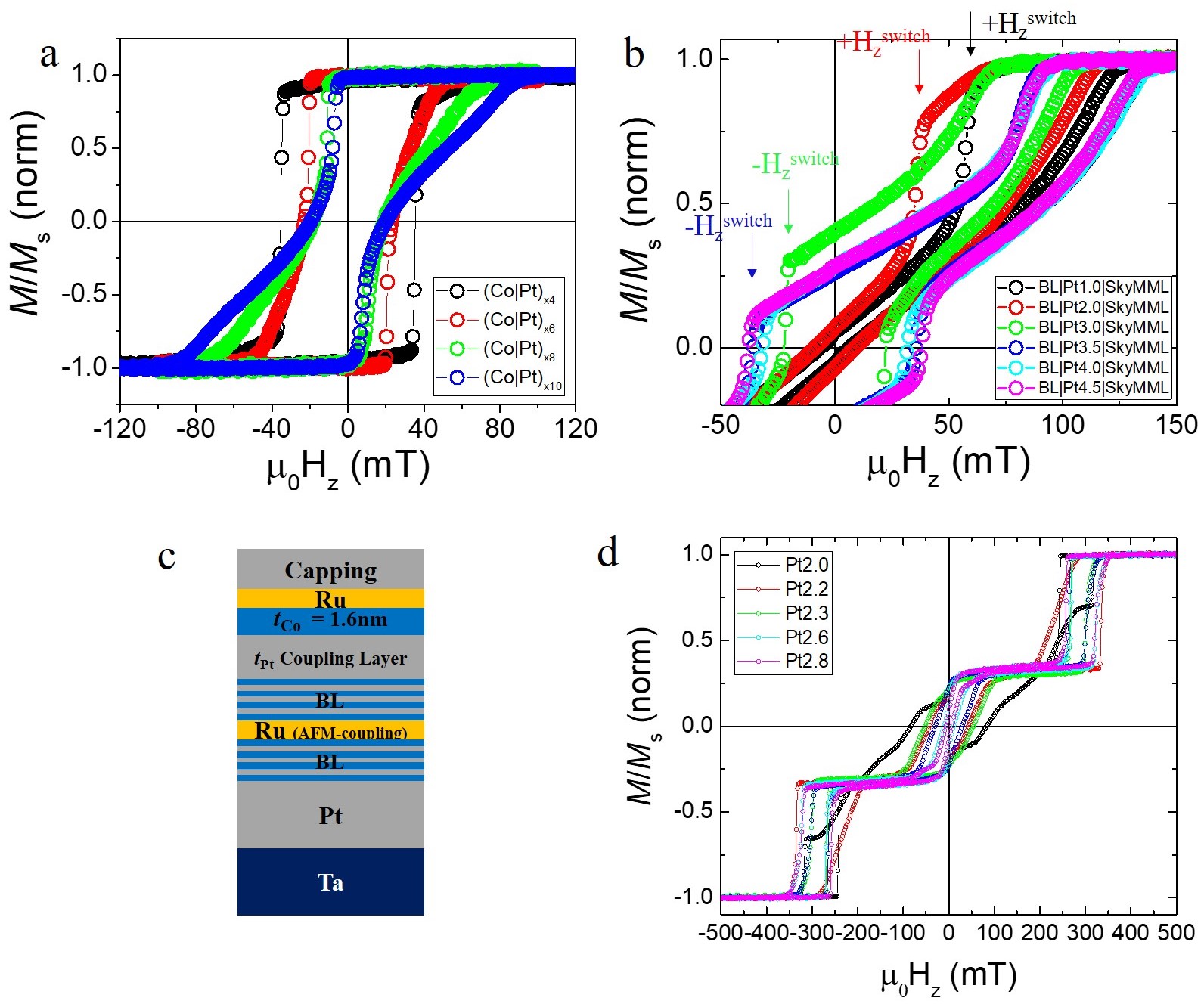}
	\caption{Bias Layer and Pt coupling layer spacer engineering. (a) AGFM hysteresis loops of BL with different number of repetitions. (b) AGFM hysteresis loops of  different Pt coupling layer thickness. The arrows point the BL $H_z$ switching. (c) Sketch of experimental samples used for extracting $H_{\rm eff}$. (d) Hysteresis loops of experimental samples represented in panel (c) varying Pt coupling layer thickness. From the satration of the central loop $H_{\rm eff}$ is estimated.  
	}
	\label{figS4}
\end{figure*}

\end{document}